\documentclass{emulateapj}

\shorttitle{Faraday rotation and polarization gradients in the jet of 3C~120}
\shortauthors{G\'omez et al.}

\begin{document}

\title{Faraday rotation and polarization gradients in the jet of 3C~120: Interaction with the external medium and a helical magnetic field?}

\author{Jos\'e L. G\'omez\altaffilmark{1}, Alan P. Marscher\altaffilmark{2}, Svetlana G. Jorstad\altaffilmark{2}, Iv\'an Agudo\altaffilmark{1} and Mar Roca-Sogorb\altaffilmark{1}}

\altaffiltext{1}{Instituto de Astrof\'{\i}sica de Andaluc\'{\i}a, CSIC, Apartado 3004, 18080 Granada, Spain. jlgomez@iaa.es; iagudo@iaa.es; mroca@iaa.es}
\altaffiltext{2}{Institute for Astrophysical Research, Boston University, 725 Commonwealth Avenue, Boston, MA 02215, USA. marscher@bu.edu; jorstad@bu.edu}

\begin{abstract}
  We present a sequence of 12 monthly polarimetric 15, 22, and 43 GHz VLBA observations of the radio galaxy 3C~120 revealing a systematic presence of gradients in Faraday rotation and degree of polarization across and along the jet. The degree of polarization increases with distance from the core and toward the jet edges, and has an asymmetric profile in which the northern side of the jet is more highly polarized. The Faraday rotation measure is also stratified across the jet width, with larger values for the southern side. We find a localized region of high Faraday rotation measure superposed on this structure between approximately 3 and 4 mas from the core, with a peak of $\sim$ 6000 rad m$^{-2}$. Interaction of the jet with the external medium or a cloud would explain the confined region of enhanced Faraday rotation, as well as the stratification in degree of polarization and the flaring of superluminal knots when crossing this region. The data are also consistent with a helical field in a two-fluid jet model, consisting of an inner, emitting jet and a sheath containing nonrelativistic electrons. However, this helical magnetic field model cannot by itself explain the localized region of enhanced Faraday rotation. The polarization electric vectors, predominantly perpendicular to the jet axis once corrected for Faraday rotation, require a dominant component parallel to the jet axis (in the frame of the emitting plasma) for the magnetic field in the emitting region.

\end{abstract}


\section{Introduction}

  It is still largely unknown what role the magnetic field plays in the dynamics and emission of relativistic jets in active galactic nuclei (AGN). Helical magnetic fields may appear naturally through the rotation of the accretion disk from which jets are launched, and could have an important role in the actual formation and collimation processes \citep{2002Sci...295.1688K,2007MNRAS..380..51K,2008Natur.452..966M}. It is possible to search for helical magnetic fields by looking for Faraday rotation measure (RM) gradients across the jet. These should appear due to the systematic change in the net line-of-sight magnetic field component across the jet, with increasing values toward the jet boundaries \citep*[e.g.,][]{Blandford:1993fk}.

  Faraday rotation gradients have been observed across the jet in 3C~273 by \citet{2002PASJ...54L..39A,2008ApJ...675...79A}, \citet{2005ApJ...626L..73Z}, and \citet{2005ApJ...633L..85A}, who interpret them as indicating the existence of a helical magnetic field, and/or the result of interaction of the jet with the ambient medium. Although transverse RM gradients have been found in other sources \citep[e.g.,][]{2004MNRAS.351L..89G}, they do not seem to be a universal feature: they are not seen in some other AGN with resolved jets \citep{2003ApJ...589..126Z}.

  The jet in the radio galaxy \object[3C120]{3C~120} is a good candidate to search for possible RM gradients across the jet. Thanks to its proximity ($z=0.033$), very long baseline interferometric observations are capable of resolving the jet across its width, revealing a very rich and dynamic structure in total and polarized flux \citep{1998ApJ...499..221G,1999ApJ...521L..29G,2000Sci...289.2317G,2001ApJ...561L.161G,2001ApJ...556..756W,2005AJ....130.1418J,2002Natur.417..625M,2007ApJ...665..232M}. Polarimetric Very Long Baseline Array (VLBA) observations of 3C~120 at 22 and 43 GHz feature differential rotation of the EVPAs of superluminal knots suggestive of Faraday rotation \citep{2000Sci...289.2317G}. In addition, evidence for the presence of a helical magnetic field has been found in 3C~120 by analyzing the motion and polarization of the knots \citep{2001ApJ...561L.161G,2005ApJ...620..646H}.

\section{Observations and data reduction}

\begin{figure}
\epsscale{1.04}
\plotone{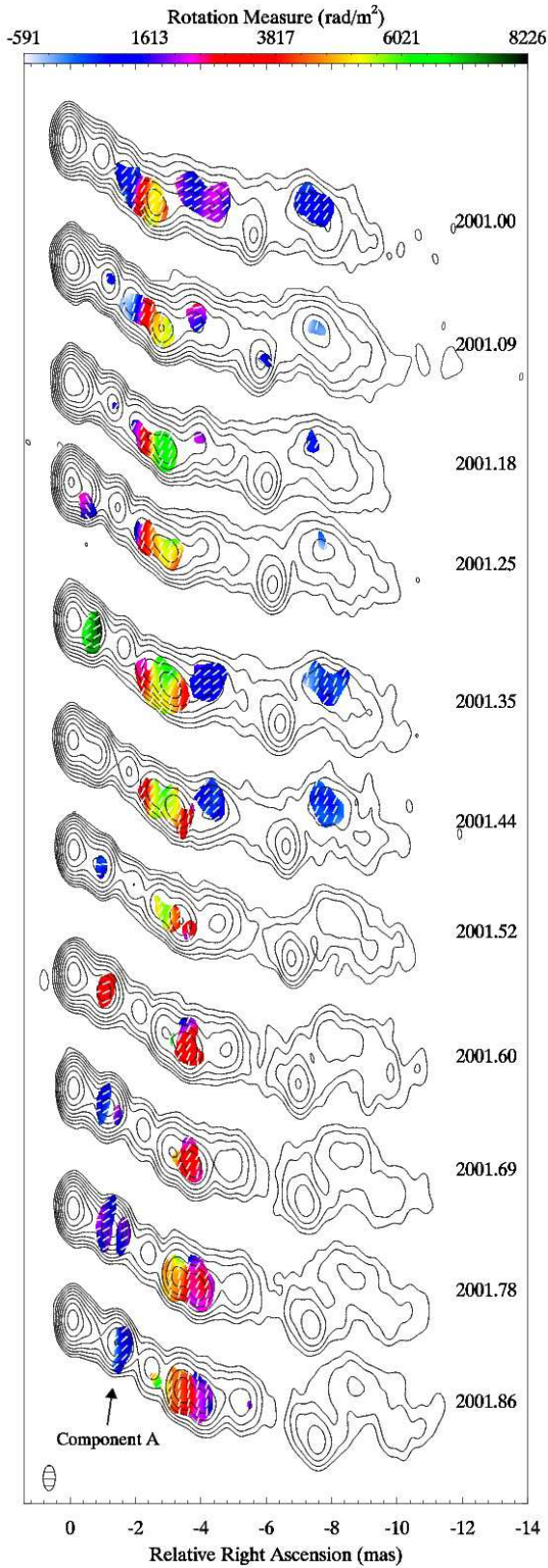}
\caption{Rotation measure maps of the jet in 3C~120. Vertical map separation is proportional to the time difference between successive epochs of observation. Total intensity (naturally weighted) 22 GHz contours are overlaid at 1.7, 3.6, 7.7, 16, 35, 74, 157, 334, and 710 mJy beam$^{-1}$. A common convolving beam of 0.78$\times$0.37 mas at -0.9$^{\circ}$ was used for all images, and is shown in the lower left corner. Bars (of unit length) indicate the RM-corrected electric vector position angle.}
\label{rm0}
\end{figure}

  The observations were made with the 10 antennas of the VLBA at the standard frequencies of 15, 22, and 43 GHz, covering a total of 12 epochs: 2000 December 30 (hereafter epoch A), 2001 February 1 (B), 2001 March 5 (C), 2001 April 1 (D), 2001 May 7 (E), 2001 June 8 (F), 2001 July 7 (G), 2001 August 9 (H), 2001 September 9 (I), 2001 October 11 (J), 2001 November 10 (K), and 2001 December 13 (L). The data were recorded in 2-bit sampling VLBA format with 32 MHz bandwidth per circular polarization and a total recording bit rate of 256 Mbits/s, except for epochs B, C, D, F, and G, for which 1-bit sampling and a total recording rate of 128 Mbits/s were used. One of the Very Large Array (VLA) antennas was added to the array at epochs A and E. Reduction of the data was performed with the AIPS software in the usual manner \citep*[e.g.,][]{1995AJ....110.2479L}. The absolute phase offset between the right- and left-circularly polarized data, which determines the electric vector position angle (EVPA), was obtained by comparison of the integrated polarization of the VLBA images of several calibrators (0420$-$014, OJ~287, BL~Lac, and 3C~454.3) with VLA observations at epochs 2000 December 31, 2001 February 3, 2001 March 4, 2001 April 6, 2001 May 11, 2001 June 10, 2001 August 12, 2001 September 10, 2001 October 12, 2001 November 6, and 2001 December 15. Estimated errors in the orientation of the EVPAs vary from epoch to epoch, but usually lie in the range of 5$^{\circ}$-7$^{\circ}$. Comparison of the D-terms across epochs provides an alternative calibration of the EVPAs \citep{2002.VLBA.SM.30}, which was found to be consistent with that obtained by comparison with the VLA data, with discrepancies smaller than 10$^{\circ}$ for most of the epochs and frequencies.

\begin{figure}
\epsscale{1.15}
\plotone{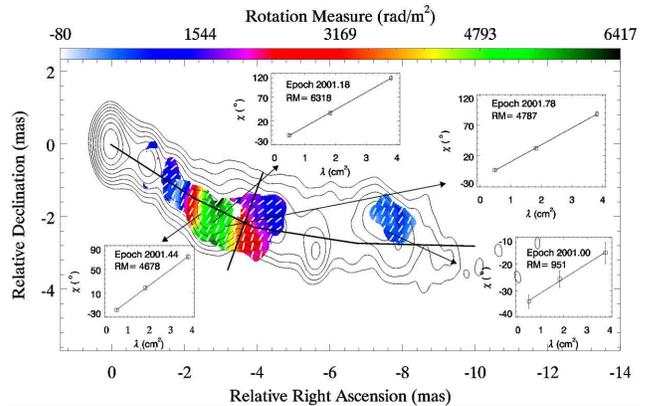}
\caption{Map of the mean value of the rotation measure maps of Fig.~\ref{rm0}. Data with standard deviation larger than 1000 rad m$^{-2}$ were discarded, which blanked most of the pixels associated with component \emph{A} while retaining most (98.4\% of the pixels) of the remaining jet. Inset panels show sample fits to a $\lambda^2$ law of the EVPAs at some particular locations and epochs. Bars indicate the mean value of the RM-corrected EVPAs, with all displayed pixels having a standard deviation smaller than 30$^{\circ}$ (96\% under 20$^{\circ}$). Contours show the 22 GHz total intensity at epoch 2001.00 for reference. The thick lines indicate the direction of the slices shown in Fig.~\ref{slices}.}
\label{rm_mean}
\end{figure}

\section{Faraday rotation and polarization gradients along and across the jet}

We have obtained rotation measure maps at each epoch from the EVPA maps, blanking the pixels for which polarization was not detected at all three frequencies simultaneously. Data at 43 and 22 GHz were first tapered and convolved with a common restoring beam to match the 15 GHz resolution. Figure \ref{rm0} presents the sequence of RM maps. (No polarization was detected at 43 GHz for epoch L owing to bad weather at some antennas.) We obtain excellent fits to a $\lambda^2$ law of the EVPAs throughout the jet, which requires $\pi/2$ rotations caused by opacity at the lower frequencies during early stages in the evolution of one new superluminal component, which is located $\sim$1.5 mas from the core at epoch 2001.86 (hereafter referred to as component \emph{A}). We note that, because of the $\pm\pi$ ambiguity of the EVPAs and the uncertainties in the estimation of the opacity, equally good fits to a $\lambda^2$ law can be found for component \emph{A} with significantly higher values of the RM (G\'omez et al., in preparation).

  The RM images of Fig.~\ref{rm0} show a changing, although very consistent, behavior across epochs. This time variability may explain the smaller RM observed by \citet{2002ApJ...566L...9Z}. Multiple superluminal components, with proper motions of the order of $\sim$ 2 mas yr$^{-1}$, sample the RM throughout the jet as they move out from the core in this sequence of 11 monthly images. This allows derivation of the rotation measure image of Fig.~\ref{rm_mean} by computing the mean value of the Fig.~\ref{rm0} maps at each pixel. Figure \ref{rm_mean} reveals a localized region of enhanced rotation measure at a distance from the core of $\sim$ 3 mas, with a peak of $\sim$ 6000 rad m$^{-2}$.

\begin{figure}
\epsscale{1.18}
\plotone{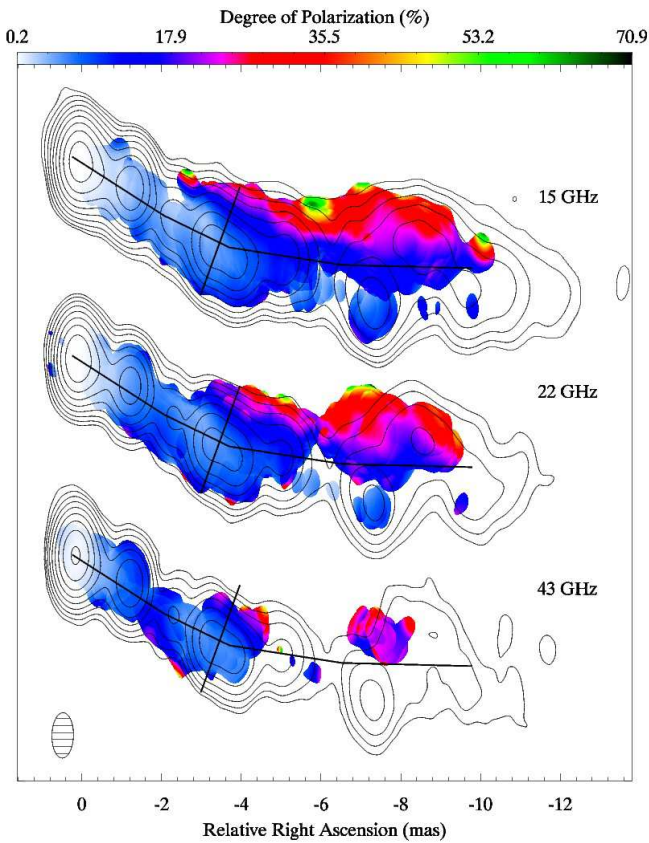}
\caption{Maps of the mean value of the degree of polarization at 15, 22, and 43 GHz obtained by combining the data at all observed epochs. Maximum standard deviations of 21\%, 20\%, and 19\%, are found, with 97\%, 98\%, and 96\% of the pixels having standard deviation smaller than 10\% at 15, 22, and 43 GHz, respectively. Data at 22 and 43 GHz were tapered and convolved with a common restoring beam of 1.14$\times$0.54 mas at position angle $-1.3^{\circ}$ (shown in the lower left corner) to match the 15 GHz resolution. Total intensity contours (epoch 2001.86) are overlaid at 0.68, 1.4, 3.0, 6.3, 13, 28, 59, 125, 263, and 556 mJy beam$^{-1}$. The thick lines indicate the direction of the slices shown in Fig.~\ref{slices}.}
\label{deg_pol}
\end{figure}

\begin{figure*}
\epsscale{1.17}
\plottwo{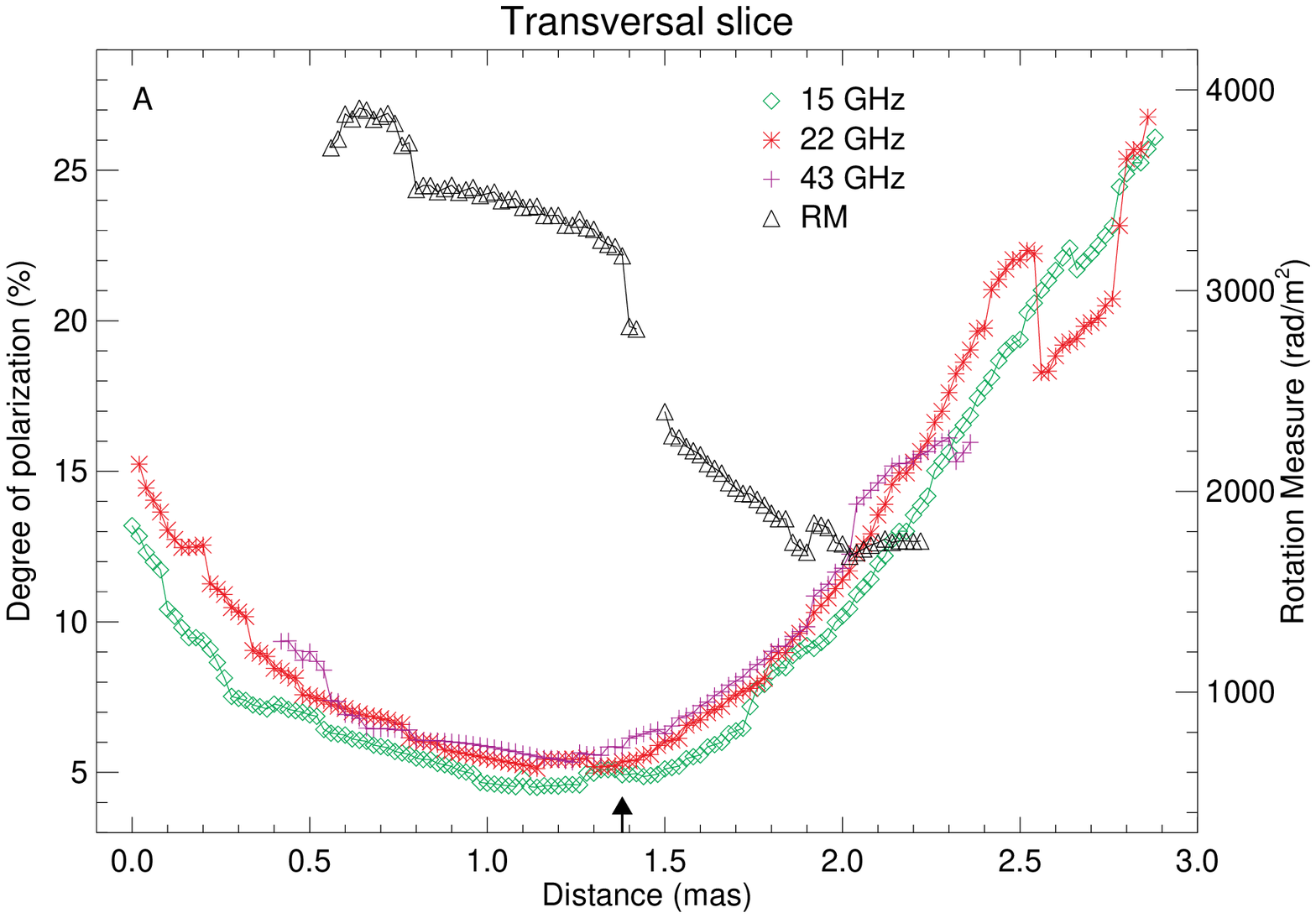}{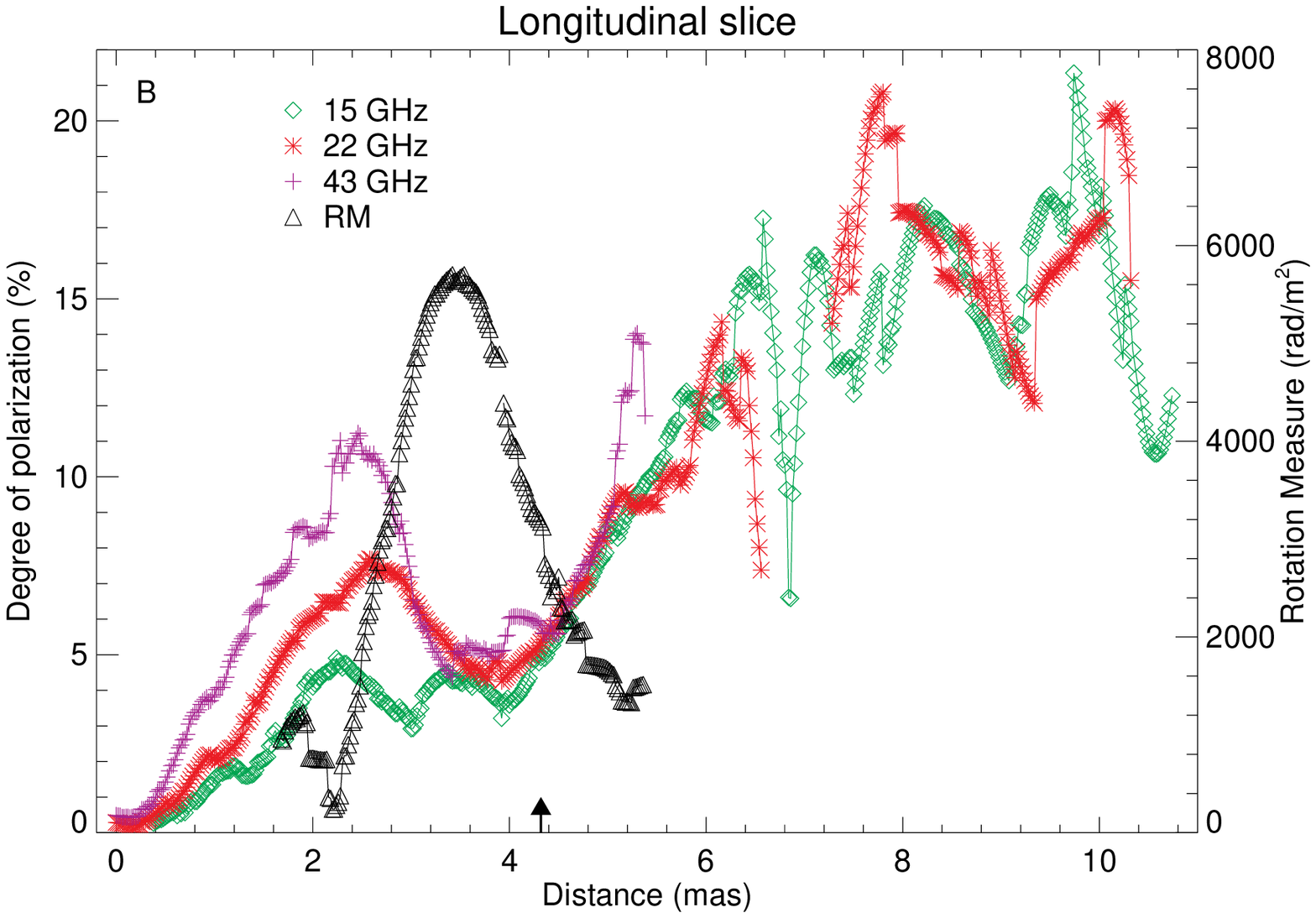}
\caption{Degree of polarization and rotation measure slices across \emph{(a; left)} and along \emph{(b; right)} the jet, as shown by the thick lines of Figs.~\ref{rm_mean} and \ref{deg_pol}. The arrow indicates the location at which both slices intersect.}
\label{slices}
\end{figure*}

  Maps of the mean value of the degree of polarization (\emph{m}), shown in Fig.~\ref{deg_pol}, reveal a clear stratification in polarization across the jet width at all three observed frequencies, with significantly larger values on the northern side of the jet. The transverse slice of Fig.~\ref{slices}a shows that \emph{m} progressively increases toward the jet edges and has a relatively weak dependence on frequency. Transverse stratification in the rotation measure is also apparent in Fig.~\ref{slices}a, with larger values on the southern side of the jet.

  The transverse profile of \emph{m} suggests a progressive reordering of the magnetic field (as integrated along the line of sight) toward the jet edges. This is consistent with the presence of a helical magnetic field and/or a shear layer produced by the interaction of the jet with the external medium \citep{2000ApJ...528L..85A,2005MNRAS.360..869L}. A helical magnetic field could also explain the observed asymmetry in the \emph{m} profile, and is in agreement with the observed transverse profile of RM. In this case the positive gradient in RM toward the southern side of the jet would require the helical magnetic field to be oriented counter-clockwise relative to the flow direction, from which it is possible to deduce the sense of rotation of the accreting disk \citep{2008ApJ...675...79A}.

  One important question is whether the RM originates in the emitting jet or in an external screen. By comparing the transverse profiles of \emph{m} and RM (Fig.~\ref{slices}a) we find that there is not a strong dependence of \emph{m} on RM and frequency, as expected for the case of internal Faraday rotation \citep{1966MNRAS.133...67B}. On the other hand, the degree of polarization increases significantly when convolving the 22 and 43 GHz images with their own, smaller beams. This is as expected for depolarization by external differential Faraday rotation, which depends on the effective angular resolution and the scale of variability of the RM and intrinsic EVPAs. Hence, although we cannot rule out some internal Faraday depolarization, the transverse profiles of \emph{m} and RM are more consistent with an external (to the emitting region) RM screen. In this case a two-fluid model, with an internal emitting jet and a sheath of thermal electrons, both immersed in a helical magnetic field, could provide an interpretation for the observed transverse profiles of \emph{m} and RM. The RM-corrected EVPAs, predominantly perpendicular to the jet axis (Figs.~\ref{rm0} and \ref{rm_mean}), require a dominant poloidal (as measured in the frame of the jet fluid) magnetic field in the emitting region \citep{2005MNRAS.360..869L}.

  The longitudinal profile (Fig.~\ref{slices}b) shows a progressive increase in \emph{m} with distance along the jet, from an unpolarized core to 15-20\% beyond $\sim$8 mas. The low polarization and its strong dependence on frequency in the inner jet regions ($\lesssim$3 mas) may be related to opacity effects, or caused by a very high and time/space variable RM \citep*[G\'omez et al., in preparation; see also][]{2005ApJ...633L..85A,2007AJ.134.799J}. A progressive reordering of the magnetic field may be required to explain the subsequent increase in degree of polarization. This may be obtained by velocity shear, as discussed in \citet{2007AJ.134.799J}. Alternatively, if moving components represent plane-perpendicular shock waves in a predominantly axis-aligned underlying magnetic field, we may expect, depending on the shock strength, a decrease in the net ordering of the field, and therefore in the degree of polarization \citep*[e.g.,][]{1998ApJ...499..221G}. Eventually, if the shock is strong enough we may expect a rotation of the EVPA as the perpendicular component of the magnetic field becomes stronger than the aligned.
Larger values of \emph{m} would be found as the compression factor of the shock decreased with distance from the core.

\section{Interaction with the external medium}

  Figure \ref{slices}b reveals the longitudinal profile of the localized region of large RM shown in Fig.~\ref{rm_mean}. The degree of polarization has a clear dependence with RM, suggesting that the decay is produced by Faraday depolarization. The rapid decrease in \emph{m} implies that there may be some internal Faraday depolarization, but it is difficult to test without further information regarding the intrinsic values of \emph{m}. This region is coincident in location, and has similar values of the RM to those postulated by \citet{2000Sci...289.2317G} from two-frequency observations. A local process, such as interaction of the jet with the external medium or a cloud, would be required in order to explain the existence of this region, as previously suggested by \citet{2000Sci...289.2317G}. The observed Faraday rotation can originate from an ionized cloud along the line of sight that may also physically interact with the jet. Some internal Faraday depolarization could also be expected if the jet partially entrains some of the thermal material of the external medium/cloud. The shock wave produced by the interaction may explain the rapid rise in emission observed for the superluminal components when crossing this region. This cloud should also cause free-free absorption of the jet emission, from which limits on its electron density, $n_e \geq 4.6 \times 10^4$ cm$^{-3}$, and magnetic field along the line of sight, $B_{\parallel} \leq 0.4 \, \mu$G, have been obtained \citep{2000Sci...289.2317G}.

  Our observations therefore support the conclusion that interaction of the jet with the external medium causes the excess Faraday rotation, stratification in the degree of polarization across the jet, and flux density flares of superluminal knots a few mas from the core. A helical magnetic field in a two-fluid jet model can be accommodated within this scenario, but by itself cannot explain the existence of the localized Faraday rotation region. If there is in fact an underlying helical field, the observed RM-corrected EVPAs require that the poloidal component (as measured in the
frame of the jet fluid) dominates over the toroidal component in the inner emitting jet.

\acknowledgements This research has been supported by the Spanish Ministerio de Educaci\'on y Ciencia and the European Fund for Regional Development through grants AYA2004-08067-C03-03 and AYA2007-67627-C03-03, and by National Science Foundation grant AST-0406865. I.\ A. acknowledges support by an I3P contract by the Spanish Consejo Superior de Investigaciones Cient\'{\i}ficas. The VLBA is an instrument of the National Radio Astronomy Observatory, a facility of the National Science Foundation operated under cooperative agreement by Associated Universities, Inc.

{\it Facilities:} \facility{VLBA ()},\facility{VLA ()}


\begin{thebibliography}{}

\bibitem[\protect\citeauthoryear{Aloy et~al.}{Aloy
  et~al.}{2000}]{2000ApJ...528L..85A}
Aloy, M.~A., G{\'o}mez, J.~L., Ib{\'a}{\~n}ez, J.~M., Mart{\'\i}, J.~M.,  \&
  M{\"u}ller, E. 2000, ApJ, 528, L85

\bibitem[\protect\citeauthoryear{Asada et~al.}{Asada
  et~al.}{2008}]{2008ApJ...675...79A}
Asada, K., Inoue, M., Kameno, S.,  \& Nagai, H. 2008, ApJ, 675, 79

\bibitem[\protect\citeauthoryear{Asada et~al.}{Asada
  et~al.}{2002}]{2002PASJ...54L..39A}
Asada, K., Inoue, M., Uchida, Y., Kameno, S., Fujisawa, K., Iguchi, S.,  \&
  Mutoh, M. 2002, PASJ, 54, L39

\bibitem[\protect\citeauthoryear{Attridge, Wardle, \& Homan}{Attridge
  et~al.}{2005}]{2005ApJ...633L..85A}
Attridge, J.~M., Wardle, J. F.~C.,  \& Homan, D.~C. 2005, ApJ, 633, L85

\bibitem[\protect\citeauthoryear{Blandford}{Blandford}{1993}]{Blandford:1993fk}
Blandford, R.~D. 1993, in Astrophysical Jets, ed. D.~Burgarella, M.~Livio, \& C.~P. O'Dea  (Cambridge: Cambridge Univ. Press), 15

\bibitem[\protect\citeauthoryear{Burn}{Burn}{1966}]{1966MNRAS.133...67B}
Burn, B.~J. 1966, MNRAS, 133, 67

\bibitem[\protect\citeauthoryear{Gabuzda, Murray, \& Cronin}{Gabuzda
  et~al.}{2004}]{2004MNRAS.351L..89G}
Gabuzda, D.~C., Murray, {\'E}.,  \& Cronin, P. 2004, MNRAS, 351, L89

\bibitem[\protect\citeauthoryear{G{\'o}mez, Marscher, \& Alberdi}{G{\'o}mez
  et~al.}{1999}]{1999ApJ...521L..29G}
G{\'o}mez, J.~L., Marscher, A.~P.,  \& Alberdi, A. 1999, ApJ, 521, L29

\bibitem[\protect\citeauthoryear{G{\'o}mez et~al.}{G{\'o}mez
  et~al.}{2001}]{2001ApJ...561L.161G}
G{\'o}mez, J.~L., Marscher, A.~P., Alberdi, A., Jorstad, S.~G.,  \& Agudo, I.
  2001, ApJ, 561, L161

\bibitem[\protect\citeauthoryear{G{\'o}mez et~al.}{G{\'o}mez
  et~al.}{2002}]{2002.VLBA.SM.30}
G{\'o}mez, J.~L., Marscher, A.~P., Alberdi, A., Jorstad, S.~G.,  \& Agudo, I.
  2002, VLBA Scientific Memo \#30

\bibitem[\protect\citeauthoryear{G{\'o}mez et~al.}{G{\'o}mez
  et~al.}{2000}]{2000Sci...289.2317G}
G{\'o}mez, J.~L., Marscher, A.~P., Alberdi, A., Jorstad, S.~G.,  \&
  Garc{\'\i}a-Mir{\'o}, C. 2000, Science, 289, 2317

\bibitem[\protect\citeauthoryear{G{\'o}mez et~al.}{G{\'o}mez
  et~al.}{1998}]{1998ApJ...499..221G}
G{\'o}mez, J.~L., Marscher, A.~P., Alberdi, A., Mart{\'\i}, J.~M.,  \&
  Ib{\'a}{\~n}ez, J.~M. 1998, ApJ, 499, 221

\bibitem[\protect\citeauthoryear{Hardee, Walker, \& G{\'o}mez}{Hardee
  et~al.}{2005}]{2005ApJ...620..646H}
Hardee, P.~E., Walker, R.~C.,  \& G{\'o}mez, J.~L. 2005, ApJ, 620, 646

\bibitem[\protect\citeauthoryear{Jorstad et~al.}{Jorstad
  et~al.}{2005}]{2005AJ....130.1418J}
Jorstad, S.~G., et~al. 2005, AJ, 130, 1418

\bibitem[\protect\citeauthoryear{Jorstad et~al.}{Jorstad
  et~al.}{2007}]{2007AJ.134.799J}
Jorstad, S.~G., et~al. 2007, AJ, 134, 799

\bibitem[\protect\citeauthoryear{Koide et~al.}{Koide
  et~al.}{2002}]{2002Sci...295.1688K}
Koide, S., Shibata, K., Kudoh, T.,  \& Meier, D.~L. 2002, Science, 295, 1688

\bibitem[\protect\citeauthoryear{Komissarov et~al.}{Komissarov
  et~al.}{2007}]{2007MNRAS..380..51K}
Komissarov, S.~S., Barkov, M.~V., Vlahakis, N.,  \& Konigl, A. 2007, MNRAS,
  380, 51

\bibitem[\protect\citeauthoryear{Leppanen, Zensus, \& Diamond}{Leppanen
  et~al.}{1995}]{1995AJ....110.2479L}
Leppanen, K.~J., Zensus, J.~A.,  \& Diamond, P.~J. 1995, AJ, 110, 2479

\bibitem[\protect\citeauthoryear{Lyutikov, Pariev, \& Gabuzda}{Lyutikov
  et~al.}{2005}]{2005MNRAS.360..869L}
Lyutikov, M., Pariev, V.~I.,  \& Gabuzda, D.~C. 2005, MNRAS, 360, 869

\bibitem[\protect\citeauthoryear{Marscher et~al.}{Marscher
  et~al.}{2002}]{2002Natur.417..625M}
Marscher, A.~P., Jorstad, S.~G., G{\'o}mez, J.~L., Aller, M.~F.,
  Ter{\"a}sranta, H., Lister, M.~L.,  \& Stirling, A.~M. 2002, Nature, 417, 625

\bibitem[\protect\citeauthoryear{Marscher et~al.}{Marscher
  et~al.}{2007}]{2007ApJ...665..232M}
Marscher, A.~P., Jorstad, S.~G., G{\'o}mez, J.~L., McHardy, I.~M., Krichbaum,
  T.~P.,  \& Agudo, I. 2007, ApJ, 665, 232
  
\bibitem[\protect\citeauthoryear{Marscher et~al.}{Marscher
  et~al.}{2008}]{2008Natur.452..966M}
Marscher, A.~P., et~al. 2008, Nature, 452, 966

\bibitem[\protect\citeauthoryear{Walker et~al.}{Walker
  et~al.}{2001}]{2001ApJ...556..756W}
Walker, R.~C., Benson, J.~M., Unwin, S.~C., Lystrup, M.~B., Hunter, T.~R.,
  Pilbratt, G.,  \& Hardee, P.~E. 2001, ApJ, 556, 756

\bibitem[\protect\citeauthoryear{Zavala \& Taylor}{Zavala \&
  Taylor}{2002}]{2002ApJ...566L...9Z}
Zavala, R.~T.,  \& Taylor, G.~B. 2002, ApJ, 566, L9

\bibitem[\protect\citeauthoryear{Zavala \& Taylor}{Zavala \&
  Taylor}{2003}]{2003ApJ...589..126Z}
Zavala, R.~T.,  \& Taylor, G.~B. 2003, ApJ, 589, 126

\bibitem[\protect\citeauthoryear{Zavala \& Taylor}{Zavala \&
  Taylor}{2005}]{2005ApJ...626L..73Z}
Zavala, R.~T.,  \& Taylor, G.~B. 2005, ApJ, 626, L73

\end{thebibliography}
\end{document}